\documentclass[amsmath,amssymb,amsbsy,reprint,prl,preprintnumbers,showpacs,superscriptaddress]{revtex4-2}
\usepackage{graphicx,color}
\usepackage{dcolumn}
\usepackage{bm}
\usepackage{braket}
\usepackage{mathtools}
\usepackage[normalem]{ulem}
\usepackage[breaklinks,colorlinks=true,linkcolor=blue,urlcolor=blue,citecolor=blue]{hyperref}
\usepackage{times}
\usepackage{physics}
\usepackage{latexsym}
\usepackage{amsmath, amssymb}
\usepackage{mathtools}
\usepackage{multirow}

\newcommand{\be}{\begin{eqnarray}}
\newcommand{\ee}{\end{eqnarray}}

\begin{document}

\title{Logarithmic Operator-Space Entanglement in Postselected Mixed States}
\date{\today}
\author{Yoshihito Kuno} 
\affiliation{Graduate School of Engineering Science, Akita University, Akita 010-8502, Japan}


\begin{abstract}
We investigate the emergence of nontrivial spatial correlations in mixed states generated by local measurements, feedback, postselection, and decoherence.
Using the doubled-Hilbert-space formulation and local conserved quantities, we exactly map the continuous-time dynamics onto a non-Hermitian Ising model with sublattice-dependent imaginary fields.
This mapping enables us to calculate the operator-space entanglement entropy (OSEE) using the fermionic correlation matrix approach.
Remarkably, starting from the maximally mixed state, we find that the steady-state OSEE exhibits logarithmic scaling at an isolated critical point separating two area-law regimes.
This critical point coincides with the closing of the bulk imaginary quasiparticle gap.
Our findings connect non-Hermitian spectral properties to the emergence of logarithmic OSEE scaling in mixed states generated solely by local quantum operations, without coherent Hamiltonian evolution.
\end{abstract}


\maketitle
\noindent{\it Introduction.---}
Quantum measurements and conditional operations provide
a route to generating and controlling nontrivial many-body correlations \cite{Lu2022_measurement,Fisher2023}.
Since the discovery of measurement-induced entanglement transitions \cite{Li2018,Skinner2019,Fisher2023}, considerable attention has been devoted to the entanglement structure emerging under nonunitary dynamics.
In particular, monitored quantum systems and nonunitary Gaussian circuits exhibit distinct entanglement scaling regimes, including area-law, logarithmic, and volume-law behavior \cite{Fuji2020,Alberton2021,Jian2022,Kells2023,
Granet2023,Murciano2023,Poboiko2023,Su2024}.
Effective non-Hermitian Hamiltonians provide a useful framework for understanding the emergence of such critical behavior from their spectral properties
\cite{Biella2021,Gopalakrishnan2021,Turkeshi2021,Turkeshi2023,Zerba2023}. 
Meanwhile, recent studies have demonstrated that measurements, feedback, and postselection can generate nontrivial orders and criticality in mixed states \cite{verresen2022,Tantivasadakarn2022,Lu2023_feedback,ziereis2025strongtoweaksymmetrybreakingphases,KO2026}.
The operator-space entanglement entropy (OSEE) \cite{Zanardi2001,Prosen2007,Znidaric2008,Prosen_PhysRevLett,Pizorn2009,Dubail2017,Nieuwenburg2018,Wellnitz2022,Carisch2023} characterizes the spatial correlation structure of a density operator, and its subsystem-size scaling can reveal criticality in operator space.
An intriguing question is whether nontrivial spatial correlations, particularly those characterized by logarithmic OSEE scaling, can emerge from a completely uncorrelated mixed state solely through local quantum operations, without coherent Hamiltonian evolution.

In this Letter, we investigate this question by constructing an exactly solvable one-dimensional model of mixed-state dynamics combining bond measurements, conditional feedback, postselection, and local decoherence.
Using local conserved quantities in the doubled Hilbert space, we exactly map the continuous-time dynamics within the relevant symmetry sector onto a non-Hermitian Ising model with sublattice-dependent imaginary fields.
The resulting Gaussian structure enables us to calculate the OSEE using the fermionic correlation matrix approach.
Remarkably, starting from the maximally mixed state, we find that tuning the decoherence rate gives rise to an isolated critical point at which the steady-state OSEE exhibits logarithmic scaling, whereas it obeys an area law on both sides.
This critical point coincides with the closing of the bulk imaginary quasiparticle gap.
The exact mapping thus allows us to understand the emergence of nontrivial spatial correlations in mixed states generated by local quantum operations from the perspective of non-Hermitian spectral properties.

\noindent{\it Postselected mixed-state dynamics.---} 
We consider a one-dimensional spin-$1/2$ chain of length $L\in4\mathbb{Z}$ with periodic boundary conditions, where sites are labeled by $j=0,\ldots,L-1$; even and odd sites form sublattices $A$ and $B$, respectively.
Throughout this work, we use the Choi mapping \cite{Choi1975,JAMIOLKOWSKI1972} to represent mixed states and quantum channels in the doubled Hilbert space.
Under this mapping, a density matrix $\rho$ acting on $\mathcal{H}$ is represented as a vector \cite{lee2023,Lee2025} in $\mathcal{H}_{u}\otimes\mathcal{H}_{\ell}$ as $|\rho\rangle\rangle\equiv \sum_{i,j}\rho_{ji}|i\rangle_u\otimes|j\rangle_\ell=\sum_i|i\rangle_u\otimes\rho|i\rangle_\ell$,
where $u$ and $\ell$ denote the upper and lower Hilbert spaces associated with the bra and ket indices, respectively.
The Choi mapping is briefly reviewed in \cite{SM}.

To this system, on each bond $(j,j+1)$, we introduce a local quantum operation described by the Kraus operators
\begin{eqnarray}
K_{0,j}&=&\sqrt{1-2J\delta t}\,I,\nonumber\\
K_{+,j}&=&\sqrt{2J\delta t}\,X_{a(j)}\Pi^{+}_{j,j+1},\nonumber\\
K_{-,j}&=&\sqrt{2J\delta t}\,\Pi^{-}_{j,j+1},
\end{eqnarray}
where $\delta t$ is an infinitesimal time step,
$J>0$ sets the measurement rate, $\Pi^{\pm}_{j,j+1}=(I\pm Z_jZ_{j+1})/2$, and $a(j)$ denotes the site on sublattice $A$ within the bond $(j,j+1)$: $a(j)=j$ for even $j$ and $a(j)=j+1$ for odd $j$.
Since $(\Pi^{\pm}_{j,j+1})^2=\Pi^{\pm}_{j,j+1}$ and $\Pi^{+}_{j,j+1}+\Pi^{-}_{j,j+1}=I$, these operators satisfy the completeness relation, $\sum_{\alpha=0,\pm}K^\dagger_{\alpha,j}K_{\alpha,j}=I$.
The physical meaning of this channel is a bond measurement of $Z_jZ_{j+1}$ accompanied by conditional feedback.
For the measurement outcome $Z_jZ_{j+1}=+1$, the feedback operation $X_{a(j)}$ flips the spin on sublattice $A$, whereas no feedback is applied for the outcome $Z_jZ_{j+1}=-1$. 
Thus, both measurement outcomes drive the bond into the sector $Z_jZ_{j+1}=-1$, while $K_{0,j}$ represents the no-measurement process. 

We also introduce an $X$-decoherence channel with rate $g_B\geq0$ acting only on sublattice $B$, $K^{(B)}_{0,j}=\sqrt{1-g_B\delta t}\,I,\:\:
K^{(B)}_{1,j}=\sqrt{g_B\delta t}\,X_j,
\quad j\in B$. These Kraus operators satisfy the completeness relation $\sum_{\alpha=0,1}K^{(B)\dagger}_{\alpha,j}K^{(B)}_{\alpha,j}=I$.

Starting from an initial density matrix $\rho(0)$, we consider a time-evolution protocol with postselection.
In what follows, we describe this dynamics using the doubled-Hilbert-space formulation, with the initial state represented by $|\rho(0)\rangle\rangle$.
At each infinitesimal time step $\delta t$, we apply the above local operations to the system.
For each bond, we perform postselection by retaining the outcomes associated with $K_{0,j}$ and $K_{+,j}$ and discarding the outcome associated with $K_{-,j}$.
We then construct a postselected mixed state by averaging over the retained measurement records with their corresponding Born probabilities, conditioned on successful postselection.
The $X$-decoherence channel on sublattice $B$ is applied without postselection.
In the following linear evolution, $\rho(t)$ denotes the unnormalized postselected density matrix.
The normalized postselected density matrix is obtained by dividing it by its trace.
At each time step $\delta t$, we first apply the bond operations sequentially in ascending order of $j$ from $0$ to $L-1$, followed by the $X$-decoherence channels on sublattice $B$. The order of these operations is irrelevant in the
continuous-time limit $\delta t\to0$. Using the doubled-space representation of the Kraus
operators \cite{lee2023,Lee2025} (see ~\cite{SM}), 
we obtain the time evolution over $\delta t$ as
\begin{eqnarray}
|\rho(t+\delta t)\rangle\rangle
=\prod_{j\in B}\mathcal{M}^{(B)}_j(\delta t)
\prod_j\mathcal{M}_j(\delta t)|\rho(t)\rangle\rangle,
\end{eqnarray}
where $\mathcal{M}_j(\delta t)
=K^*_{0,j}\otimes K_{0,j}
+K^*_{+,j}\otimes K_{+,j}
=(1-2J\delta t)I
+2J\delta t X_{a(j),u}X_{a(j),\ell}
\Pi^+_{j,j+1,u}\Pi^+_{j,j+1,\ell}$ 
and $\mathcal{M}^{(B)}_j(\delta t)
=K^{(B)*}_{0,j}\otimes K^{(B)}_{0,j}
+K^{(B)*}_{1,j}\otimes K^{(B)}_{1,j}
=(1-g_B\delta t)I
+g_B\delta t X_{j,u}X_{j,\ell}$. 
By repeating the combined operation $N_t=t/\delta t$ times, we obtain
\begin{eqnarray}
|\rho(t)\rangle\rangle
&=&\left[
\prod_{j\in B}\mathcal{M}^{(B)}_j(\delta t)
\prod_j\mathcal{M}_j(\delta t)
\right]^{N_t}|\rho(0)\rangle\rangle\nonumber\\
&\simeq&e^{-iH_{\rm NH}t}|\rho(0)\rangle\rangle,
\end{eqnarray}
where the effective non-Hermitian Hamiltonian is given by
\begin{eqnarray}
H_{\rm NH}
&=&-2iJLI-ig_B\sum_{j\in B}(I-X_{j,u}X_{j,\ell})
\nonumber\\
&&+\frac{iJ}{2}\sum_j X_{a(j),u}X_{a(j),\ell}
(I+Z_{j,u}Z_{j+1,u})\nonumber\\
&&\:\:\:\times(I+Z_{j,\ell}Z_{j+1,\ell}).
\end{eqnarray}
The exponential representation becomes exact in the continuous-time limit $\delta t\to0$.
Thus, the postselected mixed-state dynamics generated by local quantum operations is governed by the non-Hermitian Hamiltonian $H_{\rm NH}$ in the doubled Hilbert space.

\noindent{\it Exact mapping to a non-Hermitian Ising model.---}
To further analyze the non-Hermitian Hamiltonian $H_{\rm NH}$ derived above, we exploit its local conserved quantities
\cite{ziereis2025strongtoweaksymmetrybreakingphases}.
These quantities allow us to restrict the doubled-space dynamics to the sector specified by the initial state, reducing the problem to an effective spin-$1/2$ chain.

We start by introducing a local operator $Q_m=Z_{m,u}Z_{m,\ell}$ on each rung \cite{ziereis2025strongtoweaksymmetrybreakingphases}, satisfying $Q_m^2=I$ and $[Q_m,Q_n]=0$.
These operators also satisfy the commutation relations:
$[Q_m,X_{j,u}X_{j,\ell}]=0$, 
$[Q_m,Z_{j,u}Z_{j+1,u}]=[Q_m,Z_{j,\ell}Z_{j+1,\ell}]=0$. 
Thus, $[Q_m,H_{\rm NH}]=0$ for every $m$; that is, $Q_m$ is a local conserved quantity with eigenvalue $\pm 1$. 
The simultaneous eigensectors of the $Q_m$ are dynamically invariant.
For an initial density matrix diagonal in the $Z$ basis, the dynamics is restricted to the sector $Q_m=+1$ for all $m$ \cite{physical_sector_note}.
Let $\{|0\rangle_{j,s},|1\rangle_{j,s}\}$ be the eigenbasis of $Z_{j,s}$ with eigenvalues $+1$ and $-1$, respectively, where $s=u,\ell$.
In the sector $Q_j=+1$ for all $j$, the local Hilbert space on each rung is restricted to the two-dimensional subspace spanned by $\{|00\rangle_{j,u,\ell},|11\rangle_{j,u,\ell}\}$,
where $|ab\rangle_{j,u,\ell}\equiv |a\rangle_{j,u}\otimes|b\rangle_{j,\ell}$.
This reduces the dimension of the doubled Hilbert space from $4^L$ to $2^L$.
Defining the local basis states $|\tilde{0}\rangle_j=|00\rangle_{j,u,\ell}$ and $|\tilde{1}\rangle_j=|11\rangle_{j,u,\ell}$, we introduce effective Pauli operators in the reduced Hilbert space as $X_{j,u}X_{j,\ell}\mapsto\tilde{X}_j$ and $Z_{j,u},Z_{j,\ell}\mapsto\tilde{Z}_j$. 
These effective operators satisfy the standard Pauli algebra. 
By restricting $H_{\rm NH}$ to the sector $Q_j=+1$ for all $j$ and rewriting it in terms of the effective Pauli operators introduced above, we obtain
\begin{eqnarray}
\tilde{H}_{\rm NH}
&=&-i(2J+g_B/2)LI
+2iJ\sum_{j\in A}\tilde{X}_j
\nonumber\\
&&+J\sum_{j\in A}\tilde{Y}_j
(\tilde{Z}_{j-1}+\tilde{Z}_{j+1})
+ig_B\sum_{j\in B}\tilde{X}_j.
\end{eqnarray}
Here, we have used the relations $\sum_{j\in B}\tilde{X}_{j+1}=\sum_{j\in A}\tilde{X}_j$ and $\sum_{j\in B}\tilde{Z}_j\tilde{Y}_{j+1}=\sum_{j\in A}\tilde{Y}_j\tilde{Z}_{j-1}$ under periodic boundary conditions. 

To further simplify $\tilde{H}_{\rm NH}$, we introduce sublattice-dependent local unitary transformations,
$U_{A,j}=\exp[-i\pi(\tilde{X}_j+\tilde{Y}_j+\tilde{Z}_j)/(3\sqrt{3})]$ for $j\in A$ and $U_{B,j}=(\tilde{X}_j+\tilde{Z}_j)/\sqrt{2}$ for $j\in B$.
Under the local transformations $O_j\mapsto U_{\alpha,j}^\dagger O_j U_{\alpha,j}$ for $j\in\alpha$ ($\alpha=A,B$), the Pauli operators transform as
$(\tilde{X}_j,\tilde{Y}_j,\tilde{Z}_j)\mapsto(\bar{Z}_j,\bar{X}_j,\bar{Y}_j)$ on sublattice $A$
and $(\tilde{X}_j,\tilde{Y}_j,\tilde{Z}_j)\mapsto(\bar{Z}_j,-\bar{Y}_j,\bar{X}_j)$ on sublattice $B$.
These transformations correspond to a cyclic permutation of the Pauli axes on sublattice $A$ and a Hadamard transformation on sublattice $B$.
Applying these local unitary transformations to
$\tilde{H}_{\rm NH}$ and omitting the scalar term $-i(2J+g_B/2)LI$, we obtain
\begin{eqnarray}
\bar{H}_{\rm NH}
&=&J\sum^{L-1}_{j=0}\bar{X}_j\bar{X}_{j+1}
+ig_A\sum_{j\in A}\bar{Z}_j
+ig_B\sum_{j\in B}\bar{Z}_j,
\end{eqnarray}
where $g_A=2J$.  
Thus, the postselected mixed-state dynamics is exactly described by the normalized nonunitary time evolution generated by $\bar{H}_{\rm NH}$ in the reduced and rotated Hilbert space,
\begin{eqnarray}
|\rho(t)\rangle\rangle
=\frac{e^{-i\bar{H}_{\rm NH}t}|\rho(0)\rangle\rangle}
{\|e^{-i\bar{H}_{\rm NH}t}|\rho(0)\rangle\rangle\|}.
\end{eqnarray}
Hereafter, $|\rho(t)\rangle\rangle$ denotes the normalized state in this effective representation.
Here, we normalize the Choi state with respect to its Hilbert-space norm, $\sqrt{\mathrm{Tr}[\rho^2]}$ \cite{SM}, rather than the trace of the original density matrix.
Thus, in the continuous-time limit, the postselected mixed-state dynamics is exactly mapped onto the non-Hermitian Hamiltonian dynamics of an Ising chain.
Related non-Hermitian Ising models have been extensively studied in the context of monitored quantum systems \cite{Biella2021,Turkeshi2021,Turkeshi2023,Su2024}.
Importantly, in our case, the non-Hermitian dynamics
emerges in the doubled Hilbert space without coherent
Hamiltonian evolution in the original physical system.

We now specify the initial state for the postselected dynamics.
To investigate the emergence of operator-space criticality from a completely uncorrelated state, we choose the maximally mixed state $\rho(0)=I/2^L$ in the original physical system.
The corresponding normalized Choi state is given by
$|\rho(0)\rangle\rangle=\bigotimes_{j=0}^{L-1}
(|00\rangle_{j,u,\ell}+|11\rangle_{j,u,\ell})/\sqrt{2}$.
Each local rung state is an eigenstate of $\tilde X_j$
with eigenvalue $+1$.
Since $\tilde X_j$ is mapped to $\bar Z_j$ on both sublattices, the initial state in the effective spin representation is given by $|\rho(0)\rangle\rangle=|\uparrow\rangle^{\otimes L}$
in the $\bar Z_j$ basis \cite{initial_state_note}, where $\bar Z_j|\uparrow\rangle_j=|\uparrow\rangle_j$.
Starting from this product state, we investigate the steady-state correlations and their dependence on $g_B/J$.

\noindent{\it Non-Hermitian Gaussian dynamics and OSEE.---} To investigate the postselected mixed-state dynamics, we analyze the nonunitary time evolution of the effective spin chain governed by $\bar{H}_{\rm NH}$.
By employing the Jordan--Wigner transformation
\cite{Lieb1961,Mbeng2024}, we map $\bar{H}_{\rm NH}$
onto a quadratic fermionic Hamiltonian.
Since the all-up initial state in the $\bar Z_j$ basis corresponds to the fermionic vacuum \cite{fermion_vacuum_note}, the normalized nonunitary time evolution preserves the Gaussian nature of the state.
Consequently, we can analyze the dynamics and entanglement properties using the fermionic correlation matrix approach \cite{Peschel2003,PeschelEisler2009,Piccitto2022,Turkeshi2023,LeGal2024}.

To investigate how nontrivial spatial correlations emerge from the initially uncorrelated mixed state through the postselected dynamics, we focus on the correlation structure of the resulting steady mixed states. In particular, we aim to clarify how this structure changes upon tuning the decoherence rate $g_B$.
For this purpose, we employ the operator-space entanglement entropy (OSEE) \cite{Prosen2007,Pizorn2009,Znidaric2008,Nieuwenburg2014,Nieuwenburg2018}, which is defined as the bipartite entanglement entropy of the normalized Choi state (see ~\cite{SM}). We choose $\mathcal{A}$ as a contiguous spatial interval of the ladder, including both copies of each selected rung. Since the mapping is local on each rung, the OSEE of the original mixed state is exactly given by the bipartite entanglement entropy of the effective
Gaussian state. The OSEE can therefore be evaluated in the effective representation as
$S^{\rm OS}(\mathcal{A})=-{\rm Tr}[\tilde{\rho}_{\mathcal{A}}\log\tilde{\rho}_{\mathcal{A}}]$,
where $\tilde{\rho}_{\mathcal{A}}=\mathrm{Tr}_{\bar{\mathcal{A}}}\tilde{\rho}$
and $\tilde{\rho}=|\rho(t)\rangle\rangle\langle\langle\rho(t)|$
uses the normalized effective state defined above.
We can therefore directly calculate the OSEE from the fermionic correlation matrix. 
Its subsystem-size scaling diagnoses the spatial correlations encoded in the density operator
\cite{Prosen_PhysRevLett,Pizorn2009,Wellnitz2022,Orito2025,Ando2026,YK2026}.

\noindent{\it Spectrum of the non-Hermitian Ising model with sublattice-dependent imaginary fields.---} In non-Hermitian free-fermion systems, the imaginary parts
of the quasiparticle energies govern long-time eigenmode selection \cite{Biella2021,Turkeshi2021,Su2024}.
The closing of the imaginary gap can induce nonanalyticity in the steady-state correlation matrix, leading to algebraic correlations and logarithmic entanglement scaling \cite{Turkeshi2023,Xiao2026}.
We therefore examine the opening and closing of the imaginary gap of $\bar{H}_{\rm NH}$ as $g_B/J$ is varied.

Under the Jordan--Wigner transformation $c_j=(\prod^{j-1}_{m=0}\bar{Z}_m)(\bar{X}_j+i\bar{Y}_j)/2$ and $\bar{Z}_j=1-2c_j^\dagger c_j$ \cite{Lieb1961,Mbeng2024}, fermion parity is conserved.
The dynamics from the fermionic vacuum is therefore restricted to the even-parity sector, in which the Hamiltonian is
$\bar{H}_{\rm NH}
=J\sum^{L-2}_{j=0}(
c_j^\dagger c_{j+1}+c_{j+1}^\dagger c_j
+c_j^\dagger c_{j+1}^\dagger
+c_{j+1}c_j)-2i\sum^{L-1}_{j=0}g_j c_j^\dagger c_j
+i\sum^{L-1}_{j=0}g_j-J(c_{L-1}^\dagger-c_{L-1})(c_0^\dagger+c_0),
$
where $g_j=g_A$ for $j\in A$ and $g_j=g_B$ for $j\in B$.
The boundary term is incorporated by the antiperiodic fermionic boundary condition $c_{j+L}=-c_j$.

We introduce $N=L/2$ unit cells and define
$c_{A,m}\equiv c_{2m}$ and $c_{B,m}\equiv c_{2m+1}$. 
We employ the Fourier transformation
$c_{\alpha,m}=N^{-1/2}\sum_k e^{ikm}c_{\alpha,k}$
with $\alpha=A,B$, where the momentum takes the values
$k=(2n+1)\pi/N$ ($n=-N/2,\ldots,N/2-1$)
under the antiperiodic boundary condition \cite{Mbeng2024}.
By introducing the Nambu spinor
$\Psi_k=(c_{A,k},c_{B,k},c^\dagger_{A,-k},c^\dagger_{B,-k})^T$,
we write the Hamiltonian as
$\bar{H}_{\rm NH}=\frac{1}{2}\sum_k
\Psi_k^\dagger\mathcal{H}_k\Psi_k$
up to an additive constant, where the bulk momentum-space BdG Hamiltonian is given by
$
\mathcal{H}_k=
\begin{pmatrix}
h_k & \Delta_k\\
\Delta_k^\dagger & -h^T_{-k}
\end{pmatrix},
$
with
$h_k=\begin{pmatrix}
-2ig_A & J(1+e^{-ik})\\
J(1+e^{ik}) & -2ig_B
\end{pmatrix}$
and
$\Delta_k=J\begin{pmatrix}
0 & 1-e^{-ik}\\
-(1-e^{ik}) & 0
\end{pmatrix}$.
Here, $h^T_{-k}=h_k$ and $\Delta^T_{-k}=-\Delta_k$. 
From $\mathcal{H}_k$, we obtain four complex quasiparticle energy bands $\varepsilon_{\eta,\lambda}(k)=\lambda\sqrt{E_\eta^2(k)}$ with $\eta,\lambda=\pm1$, where $E_\eta^2(k)=2(2J^2-g_A^2-g_B^2)+2\eta\sqrt{D_k}$
and $D_k=(g_A^2-g_B^2)^2-4J^2(g_A^2+g_B^2+2g_Ag_B\cos k)$.
Here, $\eta$ labels the two branches of the squared spectrum, while $\lambda$ distinguishes the two eigenvalues related
by particle-hole symmetry. 
\begin{figure}[t]
\begin{center}
\includegraphics[width=7.5cm]{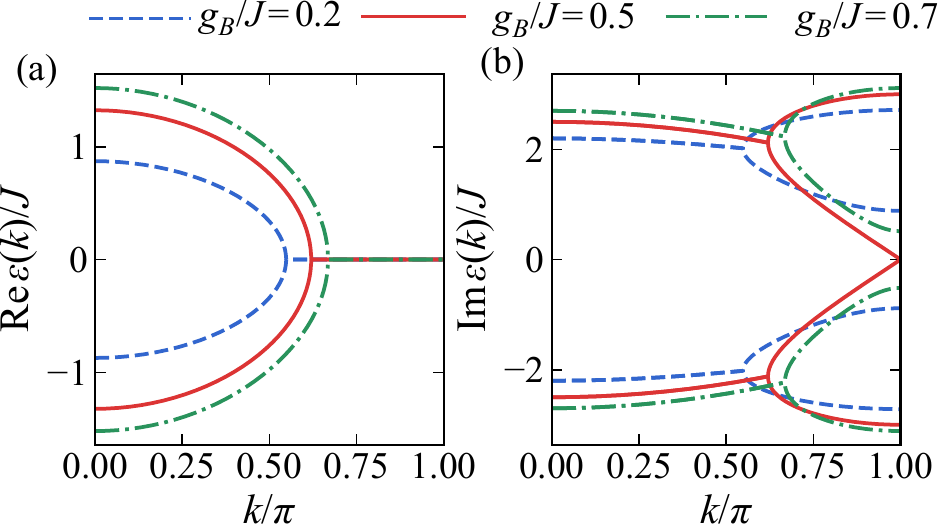}
\end{center}
\caption{Real (a) and imaginary (b) parts of the four quasiparticle
energy bands $\varepsilon_{\eta,\lambda}(k)/J$ for $g_B/J=0.2$, $0.5$, and $0.7$, with $g_A/J=2$ fixed.
Only the region $0\leq k\leq\pi$ is shown because the spectrum is symmetric under $k\to-k$.
In the momentum region with nonzero real parts, the four bands overlap pairwise in both panels, whereas all four real parts coincide at zero in the purely imaginary region.
The imaginary gap closes at $k=\pi$ for $g_B/J=0.5$, whereas it remains finite for the other two values of $g_B/J$.}
\label{Fig1}
\end{figure}
Figures~\ref{Fig1}(a) and \ref{Fig1}(b) show the real and imaginary parts of the quasiparticle energy spectrum, respectively, for representative values of $g_B/J$.
For all three values of $g_B/J$, the real parts of the four energy bands vanish in a finite momentum region near $k=\pi$.
In contrast, the imaginary gap closes at $k=\pi$ only at $g_B/J=0.5$ and remains finite on both sides of this point.
Previous studies of non-Hermitian free-fermion systems have linked gapped and gapless imaginary spectra to area-law and logarithmic entanglement scaling, respectively \cite{Turkeshi2023,Zerba2023,Xiao2026}.
Motivated by these observations, we expect that the steady-state OSEE obeys an area law on both sides of $g_B/J=0.5$ and exhibits logarithmic scaling at this isolated gap-closing point.
In the following, we examine this expectation by directly calculating the steady-state OSEE.

\noindent{\it OSEE for steady states.---}
We now examine the steady-state OSEE through the nonunitary time evolution in momentum space.
The translational symmetry with a two-site unit cell allows us to treat the Gaussian dynamics independently in each momentum block. 
We follow the normalized state $|\rho(t)\rangle\rangle$ in the effective representation, as detailed in Sec.~3 of the Supplemental Material~\cite{SM}.
In the Nambu representation, the initial fermionic vacuum is specified by two orbitals, $w_{k,1}(0)=(0,0,1,0)^T$ and $w_{k,2}(0)=(0,0,0,1)^T$.
Their unnormalized time evolution is given by $w_{k,a}(t)=e^{-i\mathcal{H}_kt}w_{k,a}(0)$ with $a=1,2$.
Collecting these orbitals into the $4\times2$ matrix $W_k(t)=(w_{k,1}(t),w_{k,2}(t))$, we obtain the two-dimensional subspace specifying the time-evolved Gaussian state.
Since the nonunitary evolution does not preserve orbital orthonormality, we employ the QR decomposition at each time
step $\Delta t$ \cite{Piccitto2022,Kawabata2023,LeGal2024,Leung2025},
$
e^{-i\mathcal{H}_k\Delta t}Q_k(t)
=Q_k(t+\Delta t)R_k(t)$,
starting from $Q_k(0)=W_k(0)$.
Here, $Q_k^\dagger(t)Q_k(t)=I_2$, and only $Q_k(t)$ is retained for the subsequent time evolution. 
The momentum-space correlation matrix is then constructed from the orthonormal orbitals as \cite{LeGal2024,Leung2025}
$C_k(t)\equiv
\langle\langle\rho(t)|\Psi_k\Psi_k^\dagger|\rho(t)\rangle\rangle
=I_4-Q_k(t)Q_k^\dagger(t)$.
Here, $Q_k(t)Q_k^\dagger(t)$ is the orthogonal projector onto the subspace spanned by the two evolved orbitals.
By iterating the above procedure to sufficiently long times, we obtain the steady-state correlation matrix $C_k^{\rm ss}=\lim_{t\to\infty}C_k(t)$.

From $C_k(t)$, the real-space correlation matrix at time $t$ is directly obtained as
$C_{mn}(t)=\frac{1}{N}\sum_k e^{ik(m-n)}C_k(t)$, where $m,n=0,\ldots,N-1$ label the unit cells.
The steady-state correlation matrix is given by $C_{mn}^{\rm ss}=\lim_{t\to\infty}C_{mn}(t)$. 

Here, we choose the subsystem $\mathcal{A}$ to consist of $L_{\mathcal{A}}$ consecutive unit cells. Restricting the real-space correlation matrix to these cells gives the $4L_{\mathcal{A}}\times4L_{\mathcal{A}}$ subsystem correlation matrix $C_{\mathcal{A}}=(C_{mn}^{\rm ss})_{m,n=0}^{L_{\mathcal{A}}-1}$. Its eigenvalues $\nu_\alpha$ \cite{NambuEigenvalues} give the single-particle entanglement energies $\xi_\alpha=\log[\nu_\alpha/(1-\nu_\alpha)]$, which appear in pairs $(\xi,-\xi)$.
The bipartite entanglement entropy is evaluated as
$S^{\rm OS}(\mathcal{A})
=-\sum_{\alpha=1}^{4L_{\mathcal{A}}}\nu_\alpha\log\nu_\alpha$.
By the local equivalence established above, this entropy equals the OSEE of the original steady mixed state.

\noindent{\it Numerical observation of the OSEE for steady mixed states.---}
We now examine the subsystem-size dependence of the steady-state OSEE.
We evolve the state with a time step $\Delta t=0.25/J$ up to $t_{\max}=8N/J$, where $N=L/2$ is the number of unit cells. 
We have confirmed that the OSEE has sufficiently converged to its steady-state value by $t_{\max}$ for the parameters considered here.

\begin{figure}[t]
\begin{center}
\includegraphics[width=7.5cm]{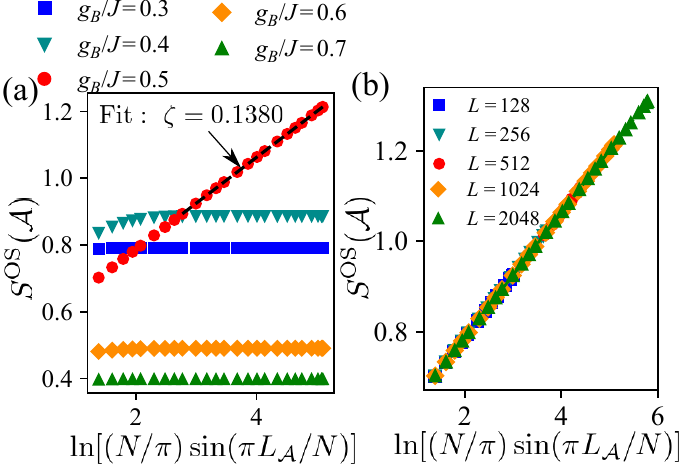}
\end{center}
\caption{Steady-state OSEE versus the logarithm of the chord length, $\log[(N/\pi)\sin(\pi L_{\mathcal{A}}/N)]$, with $g_A/J=2$ fixed. 
Here, $N=L/2$ is the total number of unit cells, and $L_{\mathcal{A}}$ is the number of consecutive unit cells in subsystem $\mathcal{A}$. 
(a) Results for $g_B/J=0.3$, $0.4$, $0.5$, $0.6$, and $0.7$ at $L=1024$ physical sites ($N=512$).
The dashed line is a logarithmic fit to the $g_B/J=0.5$ data over $16\le L_{\mathcal{A}}\le N/2$, yielding $\zeta\simeq0.1380$.
The OSEE exhibits logarithmic growth at $g_B/J=0.5$ and saturates at the four off-critical parameter values.
(b) Results at $g_B/J=0.5$ for $L=128$, $256$, $512$, $1024$, and $2048$.
The data for different system sizes nearly collapse onto a common curve, with logarithmic growth extending to larger subsystem sizes
as $L$ increases.}
\label{Fig2}
\end{figure}

Figure~\ref{Fig2}(a) compares the results of the steady OSEE for $g_B/J=0.3$, $0.4$, $0.5$, $0.6$, and $0.7$ at $L=1024$ physical sites ($N=512$). The subsystem consists of $L_{\mathcal{A}}$ consecutive unit cells,
with $4\le L_{\mathcal{A}}\le N/2$. At $g_B/J=0.5$, the OSEE grows linearly with the logarithm of the chord length and is accurately described by
$S^{\rm OS}(\mathcal{A})=\zeta\log[(N/\pi)\sin(\pi L_{\mathcal{A}}/N)]+s_0$,
where $\zeta$ and $s_0$ are fitting parameters.
A fit over $16\le L_{\mathcal{A}}\le N/2$ yields $\zeta\simeq0.1380$.
For $1\ll L_{\mathcal{A}}\ll N$, this expression reduces to $S^{\rm OS}(\mathcal{A})\simeq\zeta\log L_{\mathcal{A}}+s_0$.
In contrast, the OSEE exhibits area-law scaling for $g_B/J=0.3$, $0.4$, $0.6$, and $0.7$. Among these off-critical parameters, the saturation value is larger closer to $g_B/J=0.5$ on either side.

Figure~\ref{Fig2}(b) shows the OSEE at $g_B/J=0.5$ for
$L=128$, $256$, $512$, $1024$, and $2048$.
The data nearly collapse onto a common logarithmic curve, suggesting that the logarithmic scaling persists in the
thermodynamic limit. 
Further evidence for this critical behavior is provided by the spatial decay of the steady-state correlation matrix, which exhibits approximately algebraic decay at $g_B/J=0.5$
and a much faster decay consistent with exponential decay away from this point (see ~\cite{SM}).

Together with the spectral analysis in Fig.~\ref{Fig1}, these results identify an isolated operator-space critical point at $g_B/J=0.5$ between two area-law regimes. The logarithmic OSEE coincides with the closing of the bulk imaginary gap, whereas the area-law behavior on either side accompanies its reopening. Thus, tuning the single parameter $g_B/J$ produces an area-law--logarithmic--area-law sequence in the steady-state OSEE. Remarkably, this operator-space criticality emerges from the maximally mixed state solely through local measurements, feedback, postselection, and decoherence, without coherent Hamiltonian evolution.

\noindent{\it Conclusion.---}
We have presented an exactly solvable model of postselected mixed-state dynamics that maps onto a non-Hermitian Ising model.
Starting from the maximally mixed state, we found an isolated critical point with logarithmic OSEE scaling separating two area-law regimes.
This critical point coincides with the closing of the bulk imaginary quasiparticle gap.
Our findings demonstrate that local quantum operations alone can generate logarithmic OSEE scaling without coherent
Hamiltonian evolution.

\section*{Acknowledgements}
This work is supported by JSPS KAKENHI: 26K06956 (Y.K.).
The author used ChatGPT (OpenAI, GPT-6.2) to propose revisions to the manuscript and to assist with numerical code development and consistency checks of calculations.
The author assumes full responsibility for the content of this article.

\section*{Data availability}
The data that support the findings of this article are not publicly available. The data are available from the author upon reasonable request.
\nocite{apsrev42Control}
\bibliographystyle{apsrev4-2}
\bibliography{ref_publication_checked}
\renewcommand{\thesection}{A\arabic{section}} 
\renewcommand{\theequation}{A\arabic{equation}}
\renewcommand{\thefigure}{A\arabic{figure}}
\setcounter{equation}{0}
\setcounter{figure}{0}

\appendix
\renewcommand{\theHequation}{supp.\arabic{equation}}
\renewcommand{\theHfigure}{supp.\arabic{figure}}
\section{Supplemental Material for ``Logarithmic Operator-Space Entanglement in Postselected Mixed States''}
\section{1. Choi mapping}
Let $\rho$ be a density matrix acting on $\mathcal{H}$.
In the Choi mapping \cite{Choi1975,JAMIOLKOWSKI1972}, it is vectorized in the doubled Hilbert space $\mathcal{H}_{u}\otimes\mathcal{H}_{\ell}$ as
\begin{eqnarray}
|\rho\rangle\rangle
&\equiv&
\sum_{i,j}\rho_{ji}|i\rangle_u\otimes|j\rangle_\ell
=\sum_i|i\rangle_u\otimes\rho|i\rangle_\ell,
\end{eqnarray}
where $\rho_{ji}=\langle j|\rho|i\rangle$, and $u$ and $\ell$ denote the upper and lower Hilbert spaces associated with the bra and ket indices, respectively.
The bases $\{|i\rangle_u\}$ and $\{|j\rangle_\ell\}$ are the corresponding orthonormal bases of the two copies
of $\mathcal{H}$.
With this convention, a pure-state density matrix is mapped as 
\begin{eqnarray}
\rho_0=|\psi_0\rangle\langle\psi_0|
\quad\longrightarrow\quad
|\rho_0\rangle\rangle
=|\psi_0^*\rangle_u\otimes|\psi_0\rangle_\ell,
\end{eqnarray}
where the complex conjugation is taken in the chosen basis.
The squared norm of the Choi state is given by $\langle\langle\rho|\rho\rangle\rangle
=\Tr[\rho^\dagger\rho]=\Tr[\rho^2]$, where the last equality follows from the Hermiticity of $\rho$.

A quantum channel $\mathcal{E}$ in Kraus form is written as \cite{Nielsen2011,lidar2020}
\begin{eqnarray}
\mathcal{E}[\rho]
=\sum^{M-1}_{\alpha=0}
\hat{K}_{\alpha}\rho\hat{K}^\dagger_{\alpha},
\label{Kraus_sum}
\end{eqnarray}
where the Kraus operators satisfy the completeness relation $\sum^{M-1}_{\alpha=0}
\hat{K}^\dagger_{\alpha}\hat{K}_{\alpha}=I$.
Under the above vectorization convention, the quantum channel $\mathcal{E}$ is represented by the doubled-space operator $\hat{\mathcal{E}}$ \cite{Grover2024,Lee2025} as
\begin{eqnarray}
\mathcal{E}\longrightarrow\hat{\mathcal{E}}
=\sum^{M-1}_{\alpha=0}
\hat{K}^*_{\alpha,u}\otimes\hat{K}_{\alpha,\ell},
\end{eqnarray}
which acts as $|\mathcal{E}[\rho]\rangle\rangle=\hat{\mathcal{E}}|\rho\rangle\rangle$.
Although $\hat{\mathcal{E}}$ is generally nonunitary, the original quantum channel $\mathcal{E}$ is completely positive and trace preserving.
When postselection is performed, the sum is restricted to the Kraus operators associated with the retained outcomes, resulting in a completely positive, trace-nonincreasing map.
The normalized postselected density matrix is obtained by dividing the resulting unnormalized density matrix by its trace, provided that the trace is nonzero.

\section{2. Operator space entanglement entropy}
We briefly discuss the definition of the OSEE.
We consider a density matrix $\rho$ and its normalized Choi state, $|\tilde{\rho}\rangle\rangle=|\rho\rangle\rangle/\sqrt{\Tr[\rho^2]}$. 
The density matrix of the normalized Choi state is defined as $\tilde{\rho}\equiv
|\tilde{\rho}\rangle\rangle
\langle\langle\tilde{\rho}|$.

We partition the total system into two spatial subsystems, $\mathcal{A}$ and $\mathcal{B}$, and introduce the same spatial partition for the upper and lower Hilbert spaces as $|i\rangle_u\to|(i_{\mathcal{A}},i_{\mathcal{B}})\rangle_u$ and $|j\rangle_\ell\to|(j_{\mathcal{A}},j_{\mathcal{B}})\rangle_\ell$.
For this partition, the reduced density matrix of subsystem $\mathcal{A}$ is
\begin{eqnarray}
&&\tilde{\rho}_{\mathcal{A}}=\Tr_{\mathcal{B}}[\tilde{\rho}]\nonumber\\
&&=\sum_{i_{\mathcal{A}},i'_{\mathcal{A}},j_{\mathcal{A}},j'_{\mathcal{A}}}
\tilde{\rho}'_{\mathcal{A},(i_{\mathcal{A}},j_{\mathcal{A}}),(i'_{\mathcal{A}},j'_{\mathcal{A}})}
[|i_{\mathcal{A}}\rangle_u|j_{\mathcal{A}}\rangle_\ell
{}_{u}\langle i'_{\mathcal{A}}|{}_{\ell}\langle j'_{\mathcal{A}}|],\nonumber\\
\end{eqnarray}
where \begin{eqnarray}
&&\tilde{\rho}'_{\mathcal{A},(i_{\mathcal{A}},j_{\mathcal{A}}),(i'_{\mathcal{A}},j'_{\mathcal{A}})}
\nonumber\\
&&\equiv\sum_{i_{\mathcal{B}},j_{\mathcal{B}}}\frac{1}{\Tr[\rho^2]}
\rho_{(j_{\mathcal{A}},j_{\mathcal{B}}),(i_{\mathcal{A}},i_{\mathcal{B}})}
\rho^{*}_{(j'_{\mathcal{A}},j_{\mathcal{B}}),(i'_{\mathcal{A}},i_{\mathcal{B}})}.
\end{eqnarray}
Here, $\tilde{\rho}'_{\mathcal{A}}$ is the matrix representation of $\tilde{\rho}_{\mathcal{A}}$ in the basis
$\{|i_{\mathcal{A}}\rangle_u|j_{\mathcal{A}}\rangle_\ell\}$.

The bipartite entanglement entropy of the normalized Choi state $|\tilde{\rho}\rangle\rangle$ is therefore
given by $S^{\rm OS}(\mathcal{A})=-\Tr[\tilde{\rho}'_{\mathcal{A}}\ln\tilde{\rho}'_{\mathcal{A}}]$,
which corresponds to the OSEE of the density matrix $\rho$ \cite{Prosen2007}.
In this formulation, we take the density matrix $\rho$ as the target operator.
Similar choices of the target operator have also been considered in Refs.~\cite{Znidaric2008,Nieuwenburg2014}.

The OSEE has been widely used to characterize the spatial correlation structure and complexity of mixed-state density operators.
In particular, its system-size scaling provides a useful diagnostic of nonequilibrium phase transitions in open quantum steady states \cite{Prosen_PhysRevLett,Pizorn2009}, while more recent studies have employed it to characterize the buildup and suppression of correlations under dissipative dynamics \cite{Wellnitz2022}.

\section{3. Numerical scheme for dynamics of Gaussian states}
\label{sec:sm_gaussian_dynamics}

We describe the Gaussian evolution used in the main text. The system
contains $L\in4\mathbb{Z}$ physical sites and $N=L/2$ unit cells. 
The initial all-up state of the effective spin chain is the fermionic vacuum, whose even parity is conserved during the dynamics. 
We use antiperiodic fermionic boundary conditions, with momenta $k=(2n+1)\pi/N$, $n=-N/2,\ldots,N/2-1$, covering the full momentum grid within $-\pi<k<\pi$. 
In the Nambu representation, the initial fermionic vacuum formally corresponds to a two-particle state in a four-mode fermionic system within each Nambu block. 
The two occupied single-particle orbitals are given by
\begin{equation}
w_{k,1}(0)=
\begin{pmatrix}
0\\0\\1\\0
\end{pmatrix},
\qquad
w_{k,2}(0)=
\begin{pmatrix}
0\\0\\0\\1
\end{pmatrix}.
\end{equation}
Collecting these two orbitals, we define a $4\times2$ matrix,
\begin{equation}
W_k(0)
=\begin{pmatrix}
0&0\\
0&0\\
1&0\\
0&1
\end{pmatrix}.
\label{eq:sm3_initial_orbitals}
\end{equation}
The unnormalized evolution of these orbitals is
\begin{equation}
W_k(t)=e^{-i\mathcal H_k t}W_k(0).
\label{eq:sm3_orbital_evolution}
\end{equation}

Since the above nonunitary evolution does not preserve orbital orthonormality,
for each small discrete time step, we perform the following update by using the QR decomposition \cite{Kawabata2023,LeGal2024}
\begin{equation}
e^{-i\mathcal H_k\Delta t}Q_k(t)
=Q_k(t+\Delta t)R_k(t),
\label{eq:sm3_qr_update}
\end{equation}
starting from $Q_k(0)=W_k(0)$.
Here, $Q_k(t)$ is a $4\times2$ matrix satisfying
$Q_k^{\dagger}(t)Q_k(t)=I_2$, and $R_k(t)$ is a
$2\times2$ upper triangular matrix.

Although the full evolution of $W_k(t)$ involves the accumulated product of the matrices $R_k(t)$, each $R_k(t)$ contributes to the associated Slater determinant only through an overall factor $\det R_k$ \cite{Kawabata2023}.
Since such factors are irrelevant to the normalized Gaussian state, we retain only $Q_k(t)$ for the subsequent time evolution
and discard $R_k(t)$ at each time step
\cite{Kawabata2023,LeGal2024}.
We now show how the momentum-space correlation matrix $C_k(t)$ is obtained from the time-evolved orbital matrix $Q_k(t)$.
We write the $4\times2$ matrix $Q_k(t)$ as
\begin{equation}
Q_k(t)=
\begin{pmatrix}
q_{11}(t)&q_{12}(t)\\
q_{21}(t)&q_{22}(t)\\
q_{31}(t)&q_{32}(t)\\
q_{41}(t)&q_{42}(t)
\end{pmatrix},
\end{equation}
where $q_{ia}(t)$ is the $i$th component of the $a$th
single-particle orbital at momentum $k$.
Using these two orthonormal orbitals, we construct the normalized
two-particle wave function, whose amplitude for each ordered configuration
$(i,j)$ is given by the Slater determinant,
\begin{align}
\Phi_k(i,j;t)
&=\frac{1}{\sqrt{2}}
\begin{vmatrix}
q_{i1}(t)&q_{i2}(t)\\
q_{j1}(t)&q_{j2}(t)
\end{vmatrix}\notag\\
&=\frac{1}{\sqrt{2}}
\left[q_{i1}(t)q_{j2}(t)-q_{i2}(t)q_{j1}(t)\right],
\end{align}
where $i,j=1,\ldots,4$ label the four fermionic modes
in the Nambu representation. 
By construction, $\Phi_k(i,j;t)=-\Phi_k(j,i;t)$.
Denoting the normalized two-particle Slater state
constructed above by $|\Phi_k(t)\rangle$, we calculate
the one-particle correlation matrix as
\begin{align}
&\langle\Phi_k(t)|\Psi_{k,j}^{\dagger}\Psi_{k,i}|\Phi_k(t)\rangle=2\sum_{l=1}^{4}\Phi_k(i,l;t)\Phi_k^{*}(j,l;t)\notag\\
&=\sum_{a=1}^{2}q_{ia}(t)q_{ja}^{*}(t)=[Q_k(t)Q_k^{\dagger}(t)]_{ij},
\end{align}
where $\Psi_{k,i}$ denotes the $i$th component of the Nambu spinor defined in the main text, and we have used the orthonormality of the two orbitals,
$Q_k^{\dagger}(t)Q_k(t)=I_2$.
Using the fermionic anticommutation relations, we obtain
\begin{equation}
C_k(t)=
\langle\Phi_k(t)|\Psi_k\Psi_k^{\dagger}|\Phi_k(t)\rangle
=I_4-Q_k(t)Q_k^{\dagger}(t).
\label{eq:sm3_correlation_matrix}
\end{equation}
From the correlation matrix $C_k(t)$ obtained in this way, we calculate the real-space correlation matrix and the OSEE as described in the main text.

\section{4. Spatial decay of the correlation matrix}

\begin{figure}[b]
\begin{center}
\includegraphics[width=6.5cm]{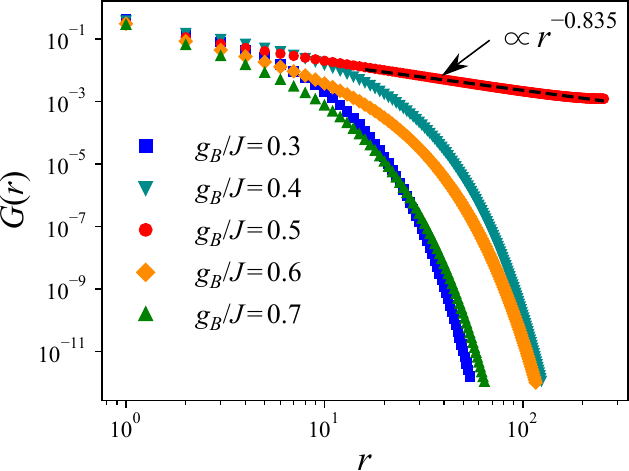}
\end{center}
\caption{Spatial decay of the Frobenius norm
$G(r)=\|C_{0,r}^{\rm ss}\|_{\rm F}$
of the steady-state correlation matrix for
$g_B/J=0.3$, $0.4$, $0.5$, $0.6$, and $0.7$,
with $g_A/J=2$ fixed and $L=1024$ physical sites
($N=512$ unit cells).
The separation $r$ is measured in unit cells,
and both axes are logarithmic.
At $g_B/J=0.5$, $G(r)$ exhibits approximately algebraic decay.
The black dashed line shows the power-law fit
$G(r)\propto r^{-0.835}$,
with the exponent characterizing the finite-distance behavior.
The four off-critical curves decay much more rapidly,
consistent with exponential decay.}
\label{FigA1}
\end{figure}

To complement the spectral and OSEE analyses in the main text,
we examine the spatial decay of the steady-state correlation matrix.
The $4\times4$ block $C_{0,r}^{\rm ss}$ contains the normal
and anomalous fermionic correlations between two unit cells
separated by $r$ in the effective Gaussian state.
We characterize their overall magnitude by the Frobenius norm,
\begin{eqnarray}
G(r)&=&\left\|C_{0,r}^{\rm ss}\right\|_{\rm F}=\sqrt{\sum_{\alpha,\beta=1}^{4}
\left|\left[C_{0,r}^{\rm ss}\right]_{\alpha\beta}\right|^2},
\end{eqnarray}
where $r>0$.
This quantity combines the contributions from all components
of the correlation block into a single nonnegative measure.

Figure~\ref{FigA1} shows $G(r)$ for
$g_B/J=0.3$, $0.4$, $0.5$, $0.6$, and $0.7$,
with $g_A/J=2$ fixed.
At $g_B/J=0.5$, the correlations decay approximately algebraically,
as illustrated by the power-law fit $G(r)\propto r^{-0.835}$.
The fitted exponent describes the finite-distance behavior
of the present system.
For the four off-critical parameters,
$G(r)$ instead exhibits pronounced downward curvature
on the double-logarithmic plot,
consistent with exponential decay and a finite correlation length.
The decay becomes slower as $g_B/J$ approaches $0.5$
from either side.

These observations are consistent with the bulk spectral
structure in Fig.~\ref{Fig1} and the OSEE scaling
in Fig.~\ref{Fig2}.
The algebraic decay occurs at the closing of the bulk imaginary gap,
where the steady-state OSEE grows logarithmically.
The rapid off-critical decay accompanies a finite imaginary gap
and area-law OSEE.
The correlation matrix thus provides complementary evidence
for the isolated operator-space critical point.
Since these fermionic correlators are evaluated in the normalized
doubled state, they characterize the correlations in its effective
Gaussian representation.

\end{document}